# Generalized theory of spatial coherence for superposition of two speckle patterns with polarization diversity

Abhijit Roy

École Polytechnique Fédérale de Lausanne (EPFL), Lausanne 1015, Switzerland

Email: abhijitphy302@gmail.com

A generalized theory of spatial coherence for superposition of two speckle patterns with polarization diversity is presented. The presented theory deals with superposition in different scenarios i.e. superposition of two fully correlated, partially correlated and completely uncorrelated speckle patterns, and describes the effect on the spatial coherence-polarization (CP) property of the superposed random field from the study of the spatial degree of coherence and degree of polarization. The change in the spatial CP property with the polarization diversity is studied for different correlation factors of the constituent random fields, and a modulation of the spatial CP property is observed, when the constituent speckle patterns are not fully correlated. The effect of the average intensity ratio of the two random fields on the spatial CP property is also studied. Thus, this theoretical study completes the case of superposition of two speckle patterns with polarization diversity and its effect on the spatial CP property.

A random intensity distribution, referred as speckle pattern, is observed due to the interference of randomly scattered light. Although analyzing a single speckle pattern has been useful for various imaging related study through a random scattering medium, the superposition of speckle patterns has recently attracted the attention due to its wide range of applications and versatility. The superposition of two uncorrelated speckle patterns can be used in off-axis speckle holography for looking through an optical barrier [1, 2]. Sensing of polarization rotation through a scattering medium and controlling the spatial polarization distribution of speckles can be achieved from the superposition of two uncorrelated speckle patterns with polarization diversity [3, 4]. Speckle interferometry is another area, where the superposition of two speckle patterns are utilized for tomographic measurement and also to study in-plane deformation [5, 6].

The concept of the superposition of speckle patterns has been widely used in different biomedical studies. Capturing a large number of temporally fluctuating speckle patterns and calculating its average is useful in eliminating temporal fluctuation from the measurement [7]. A similar approach can be adopted to eliminate the speckles present in a recorded image and has been used for imaging through a random scattering medium [8]. The contrast of a speckle pattern can be reduced, if superposition of independent speckle patterns is taken and the contrast is reported to be reducing with the increase in the number of speckle patterns [9]. A modulation of the contrast can also be achieved, if polarization diversity is incorporated in the superposition of correlated speckle patterns [10].

The importance of proper characterization of the superposition of speckle patterns has increased due to the mentioned wide range of applications. It has been reported that the probability density function of intensity of a superposed speckle pattern changes with the mutual correlation of the constituent speckle patterns [11]. The probability density function is found to be following an exponential decay behavior in case of speckle patterns, which are fully correlated and with the reduction of the mutual correlation, the function is observed to be deviating from the exponential decay nature [12]. Moreover, efforts have also been made to study the average contrast in case of superposition of mutually correlated and partially correlated speckle patterns [13, 14]. The Shannon entropy of the superposed speckle pattern has also been reported to be changing with the polarization diversity of the constituent uncorrelated speckle patterns [15].

The aforementioned reported works are mostly limited to the scalar domain i.e. polarization diversity has not been largely introduced in the case of superposition of speckle patterns. Moreover, the works, which deal with the polarization diversity in the superposed speckle pattern are confined either to the case of uncorrelated speckle patterns or fully correlated speckle patterns [3, 4, 10, 15, 16]. Here, in this work, a generalized theory is developed to study the spatial coherence-polarization (CP) property of the superposed speckle pattern in case of superposition of two speckle patterns with polarization diversity. The case of the superposition of two uncorrelated speckle patterns, as reported in Refs. [4, 16], can be derived from this generalized theory, which can also be used to study the effect on the CP property, if the mutual correlation of the constituent speckle patterns is changed.

Let us consider that the polarization vector of an object random field $\mathbf{E_O}(\mathbf{r}, t)$ makes an angle $\theta$ with the x-axis and can be written as

$$\mathbf{E_O}(\mathbf{r}, t) = E_O(\mathbf{r}, t)(\cos\theta\, \hat{\mathbf{x}} + \sin\theta\, \hat{\mathbf{y}}) \tag{1}$$

where $E_O(\mathbf{r}, t)$ is the amplitude of the object random field, $\mathbf{r}$ is the spatial position vector on the transverse observation plane, and $\hat{\mathbf{x}}$ and $\hat{\mathbf{y}}$ are the two mutual orthonormal vectors. Similarly, if a reference random field $\mathbf{E_R}(\mathbf{r}, t)$ makes an angle $\phi$ with the x-axis, it can be written as

$$\mathbf{E_R}(\mathbf{r}, t) = E_R(\mathbf{r}, t)(\cos\phi\, \hat{\mathbf{x}} + \sin\phi\, \hat{\mathbf{y}}) \tag{2}$$

The superposition of the object and reference random fields can be written as

$$\mathbf{E}(\mathbf{r}, t) = \mathbf{E_O}(\mathbf{r}, t) + \mathbf{E_R}(\mathbf{r}, t) \tag{3}$$

As mentioned earlier, the spatial CP property of the superposed random field is investigated from the study of the spatial degree of coherence (DoC), $\gamma(\mathbf{r_1}, \mathbf{r_2})$ and degree of polarization (DoP), $P(\mathbf{r})$, which can be calculated from an intensity distribution, $I(\mathbf{r})$ following the intensity correlation based approach or from the CP matrix, $\Gamma(\mathbf{r_1}, \mathbf{r_2})$ of the random field following Ref. [4] as

$$\gamma^2(\mathbf{r_1}, \mathbf{r_2}) = \frac{\langle \Delta I(\mathbf{r_1})\, \Delta I(\mathbf{r_2}) \rangle}{\langle I(\mathbf{r_1}) \rangle \langle I(\mathbf{r_2}) \rangle} = \frac{\mathrm{tr}[\Gamma(\mathbf{r_1}, \mathbf{r_2})\, \Gamma^\dagger(\mathbf{r_1}, \mathbf{r_2})]}{|\mathrm{tr}[\Gamma(0)]|^2} \tag{4}$$

$$P^2(\mathbf{r}) = 2\gamma^2(\mathbf{r}, \mathbf{r}) - 1 \tag{5}$$

where $\Delta I(\mathbf{r}) = I(\mathbf{r}) - \langle I(\mathbf{r}) \rangle$ is the deviation of intensity from its mean value, '$\langle . \rangle$' represents the ensemble average of the variable, 'tr' denotes the trace of the matrix, and $\gamma(\mathbf{r}, \mathbf{r})$ is the maximum DoC. The CP matrix, $\Gamma(\mathbf{r_1}, \mathbf{r_2})$ is a 4×4 matrix and the matrix elements are defined as $\Gamma_{ij}(\mathbf{r_1}, \mathbf{r_2}) = \langle E_i^*(\mathbf{r_1})\, E_j(\mathbf{r_2}) \rangle$. In the present study, the $\gamma^2(\mathbf{r_1}, \mathbf{r_2})$ of the superposed random field, $\mathbf{E}(\mathbf{r}, t)$ is determined from the intensity correlation based approach under the assumption that the random field follows Gaussian statistics, and it is observed that a similar expression of $\gamma^2(\mathbf{r_1}, \mathbf{r_2})$ in terms of $\Gamma(\mathbf{r_1}, \mathbf{r_2})$, as shown in Eq. (4), can be derived, where $\Gamma(\mathbf{r_1}, \mathbf{r_2})$ can be expressed as

$$\Gamma(\mathbf{r_1}, \mathbf{r_2}) = \Gamma^{OO}(\mathbf{r_1}, \mathbf{r_2}) + \Gamma^{OR}(\mathbf{r_1}, \mathbf{r_2}) + \Gamma^{RO}(\mathbf{r_1}, \mathbf{r_2}) + \Gamma^{RR}(\mathbf{r_1}, \mathbf{r_2}) \tag{6}$$

It can be observed from Eq. (6) that the CP matrix of the superposed random field can be written as a sum of the CP matrix of the object random field, reference random field and the CP matrices due to the interference of these two fields i.e. $\Gamma^{OO}(\mathbf{r_1}, \mathbf{r_2})$, $\Gamma^{RR}(\mathbf{r_1}, \mathbf{r_2})$, and $\Gamma^{OR}(\mathbf{r_1}, \mathbf{r_2})$ and $\Gamma^{RO}(\mathbf{r_1}, \mathbf{r_2})$, respectively. As the DoC at $\mathbf{r_1} = \mathbf{r_2}$ denotes the maximum DoC, which is used to calculate the DoP, the calculation is focused at $\mathbf{r_1} = \mathbf{r_2}$. After inserting Eqs. (1) and (2), the CP matrices in Eq. (6) at $\mathbf{r_1} = \mathbf{r_2}$ can be written as

$$\Gamma^{OO}(\mathbf{r}, \mathbf{r}) = \langle E_O^*(\mathbf{r})\, E_O(\mathbf{r}) \rangle \begin{pmatrix} \cos^2\theta & \sin\theta\cos\theta \\ \sin\theta\cos\theta & \sin^2\theta \end{pmatrix} \tag{7}$$

$$\Gamma^{OR}(\mathbf{r}, \mathbf{r}) = \langle E_O^*(\mathbf{r})\, E_R(\mathbf{r}) \rangle \begin{pmatrix} \cos\theta\cos\phi & \cos\theta\sin\phi \\ \sin\theta\cos\phi & \sin\theta\sin\phi \end{pmatrix} \tag{8}$$

$$\Gamma^{RO}(\mathbf{r}, \mathbf{r}) = \langle E_R^*(\mathbf{r})\, E_O(\mathbf{r}) \rangle \begin{pmatrix} \cos\phi\cos\theta & \cos\phi\sin\theta \\ \sin\phi\cos\theta & \sin\phi\sin\theta \end{pmatrix} \tag{9}$$

$$\Gamma^{RR}(\mathbf{r}, \mathbf{r}) = \langle E_R^*(\mathbf{r})\, E_R(\mathbf{r}) \rangle \begin{pmatrix} \cos^2\phi & \sin\phi\cos\phi \\ \sin\phi\cos\phi & \sin^2\phi \end{pmatrix} \tag{10}$$

The terms outside the matrix in Eqs. (7) and (10) can be written as $\langle I_O(\mathbf{r}) \rangle$ and $\langle I_R(\mathbf{r}) \rangle$, respectively, following the fact that $\langle I_i(\mathbf{r}) \rangle = \langle E_i^*(\mathbf{r})\, E_i(\mathbf{r}) \rangle$. The terms outside the matrix in Eqs. (8) and (9) denote the mutual correlation of the object and reference random fields, and can be written in terms of the degree of mutual correlation, *g* following Ref. [17] as

$$g = \frac{\langle E_O^*(\mathbf{r}) E_R(\mathbf{r})\rangle}{\sqrt{\langle I_O(\mathbf{r})\rangle \langle I_R(\mathbf{r})\rangle}} \tag{11}$$

We define another parameter, *m* which denotes the ratio of the average intensity of the object and reference random fields i.e. $m = \frac{\langle I_O(\mathbf{r})\rangle}{\langle I_R(\mathbf{r})\rangle}$. The Eqs. (7-10) can be written in terms of *m* and *g* as

$$\Gamma^{OO}(\mathbf{r},\mathbf{r}) = m \langle I_R(\mathbf{r})\rangle \begin{pmatrix} \cos^2\theta & \sin\theta\cos\theta \\ \sin\theta\cos\theta & \sin^2\theta \end{pmatrix} \tag{12}$$

$$\Gamma^{OR}(\mathbf{r},\mathbf{r}) = g\sqrt{m} \langle I_R(\mathbf{r})\rangle \begin{pmatrix} \cos\theta\cos\phi & \cos\theta\sin\phi \\ \sin\theta\cos\phi & \sin\theta\sin\phi \end{pmatrix} \tag{13}$$

$$\Gamma^{RO}(\mathbf{r},\mathbf{r}) = g\sqrt{m} \langle I_R(\mathbf{r})\rangle \begin{pmatrix} \cos\phi\cos\theta & \cos\phi\sin\theta \\ \sin\phi\cos\theta & \sin\phi\sin\theta \end{pmatrix} \tag{14}$$

$$\Gamma^{RR}(\mathbf{r},\mathbf{r}) = \langle I_R(\mathbf{r})\rangle \begin{pmatrix} \cos^2\phi & \sin\phi\cos\phi \\ \sin\phi\cos\phi & \sin^2\phi \end{pmatrix} \tag{15}$$

The Eqs. (12-15) are used in Eq. (6) to determine the $\Gamma(\mathbf{r},\mathbf{r})$ of the superposed random field. The estimated $\Gamma(\mathbf{r},\mathbf{r})$ is then used in Eq. (4) to calculate the $\gamma^2(\mathbf{r},\mathbf{r})$ of the superposed random field and can be written as

$$\gamma^2(\mathbf{r},\mathbf{r}) = \frac{N_{uc} + N_c}{(D_{uc} + D_c)^2} \tag{16}$$

where $N_{uc} = 1 + m^2 + 2m\cos^2(\theta - \phi)$ and

$N_c = 2g^2 m[1 + \cos^2(\theta - \phi)] + 4g\sqrt{m}(1+m)\cos(\theta - \phi); D_{uc} = 1 + m; D_c = 2g\sqrt{m}\cos(\theta - \phi)$.

It can be observed from Eq. (16) that the maximum DoC consists of two different types of terms both in the numerator and denominator: $N_c$ and $D_c$, which are due to the correlation between the two constituent random fields, which can be attributed due to the presence of the correlation factor *g*, and $N_{uc}$ and $D_{uc}$, which can be considered as the contribution, when the speckle patterns are not mutually correlated i.e. the value of *g* is zero. Similarly, the value of $P^2(\mathbf{r})$ is estimated from Eq. (5) using Eq. (16), and can be found as

$$P^2(\mathbf{r}) = \frac{P_{uc} + P_c}{(D_{uc} + D_c)^2} \tag{17}$$

where

$P_{uc} = 1 + m^2 + 2m\cos 2(\theta - \phi); P_c = 4g^2 m + 4g\sqrt{m}(1+m)\cos(\theta - \phi)$.

Eqs. (16) and (17) are the generalized expressions of the maximum DoC and DoP, respectively, of the superposition of two speckle patterns with polarization diversity and having different average intensities. If the two constituent speckle patterns are considered to be mutually uncorrelated i.e. $g = 0$, the expressions of the maximum DoC and DoP, as reported in Ref. [16] can be derived from Eqs. (16) and (17). Apart from $g = 0$, if the value of *m* is taken as unity i.e. $m = 1$, which indicates that the two constituent random fields have the same average intensity, it can be found that Eqs. (16) and (17) take the following form

$$\gamma^2(\mathbf{r},\mathbf{r}) = \frac{1}{2}[1 + \cos^2(\theta - \phi)] \tag{18}$$

$$P^2(\mathbf{r},\mathbf{r}) = \cos^2(\theta - \phi) \tag{19}$$

These equations have been reported and experimentally verified in Ref. [4], where the two uncorrelated speckle patterns have been generated from two statistically independent scattering media. On the other hand, if superposition of two fully correlated speckle patterns with equal average intensity is considered, it can be found from Eqs. (16) and (17) that the maximum DoC and DoP are unity and are independent of the polarization diversity of the constituent speckle patterns. This is due to the fact that as the two constituent speckle patterns are fully correlated, it means that they are same and hence, the superposed speckle pattern will have the same maximum DoC and DoP as that of the constituent speckle patterns i.e. unity. This situation can be achieved, if the two speckle patterns are generated from a single scattering layer with the incident beam propagating along the same directions in both the cases. It has been reported by Freund *et al.* that in case of a scattering layer, the generated speckle pattern does not change with the change of the polarization diversity [18]. In this case, the superposed speckle pattern has unity maximum DoC and DoP.

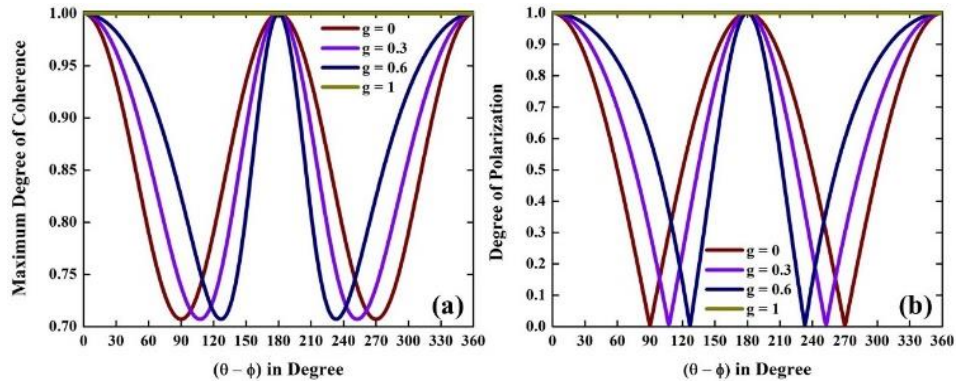

Fig. 1. The variations of the (a) maximum DoC and (b) DoP with $(\theta - \phi)$ for different values of $g$ and for $m = 1$.

In order to understand the effect of mutual correlation of two speckle patterns on the variation of the maximum DoC and DoP of the superposed speckle pattern with $(\theta - \phi)$, the value of $m$ is taken unity and the change of the maximum DoC and DoP with $(\theta - \phi)$ is studied for different values of $g$, and the variations are presented in Fig. 1. It can be observed from Fig. 1 that the value of $g$ has a tremendous effect on the spatial coherence of the superposed speckle pattern, where it is found that in case of g = 1, no modulation of the maximum DoC and DoP with $(\theta - \phi)$ are observed. On the other hand, for g = 0, a sinusoidal modulation of the maximum DoC and DoP with $(\theta - \phi)$ can be found. For other values of $g$ i.e. for $0 < g < 1$, the positions of the minima of the maximum DoC and DoP are observed to be shifting monotonically from $(\theta - \phi) = 90^0$ and $270^0$ to a greater value (for $90^0$) and to a lower value (for $270^0$), respectively, and the positions of the minima are also observed to be shifting towards each other, which makes a sharper variation of the maximum DoC and DoP for $90^0 < (\theta - \phi) < 270^0$.

The effect of the value of $m$ on the modulation of the maximum DoC and DoP with $(\theta - \phi)$ is studied for different values of $g$. As the variations of the maximum DoC and DoP are related to each other, in this study, we have focused on the variation of the DoP, and the variations for g = 0.3 and 0.6 are shown in Figs. 2(a) and 2(b), respectively. It can be observed that in both the cases, the difference between the maxima and minima of the variation reduces with the increase of the value of $m$.

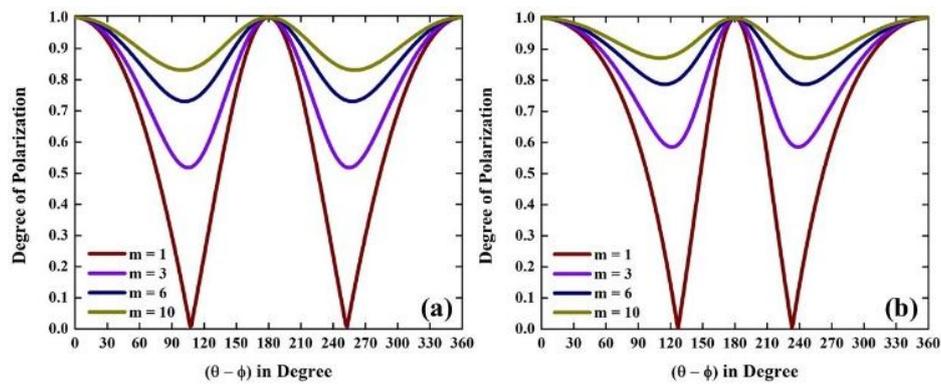

Fig. 2. The change of the DoP with $(\theta - \phi)$ for g = (a) 0.3 and (b) 0.6, for different values of $m$.

It is found from Figs. 1 and 2 that change in $m$ and $g$ have a tremendous effect on the variation of the spatial CP property with $(\theta - \phi)$. This effect is clearly observed from the change of the magnitude and position of the minimum of the DoP. A detailed study is carried out for a better understanding of these effects, where the variations of the magnitude and position of the minimum DoP is studied as functions of $g$ and $m$ for different fixed values of $m$ and $g$, respectively, and the results are presented in Figs. 3 and 4.

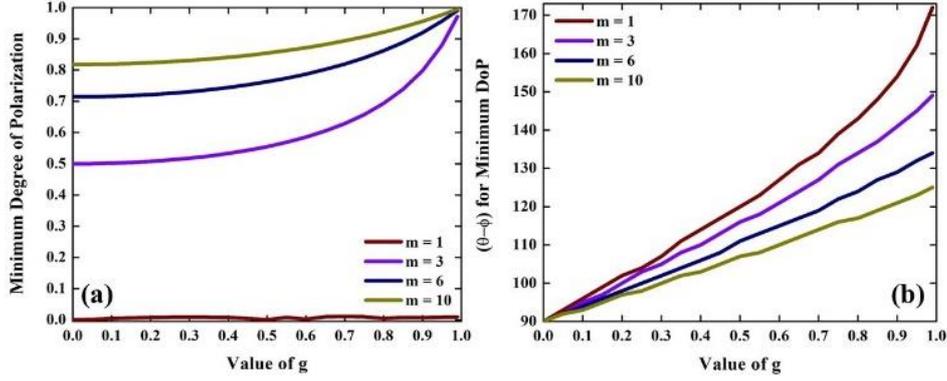

Fig. 3. Change of the (a) magnitude and (b) position of minimum DoP with *g* for different values of *m*.

In Fig. 3(a), the magnitude of the minimum DoP is found to be zero and independent of *g*, when the two constituent speckle patterns have the same average intensity i.e. m = 1. However, the position of the minimum DoP is observed to be shifting from $90^0$ to $180^0$ with the increase of *g* (Fig. 3(b)), as also seen in Fig. 1. On the other hand, with the increase of the value of *m*, the magnitude of minimum DoP is found to be monotonically increasing from the value at g = 0, and reaches very close to 1 when the value of *g* is also near to unity. In Fig. 3(b), the rate of change in the position of the minimum DoP with *g* is found to be reducing with the increase of the value of *m*.

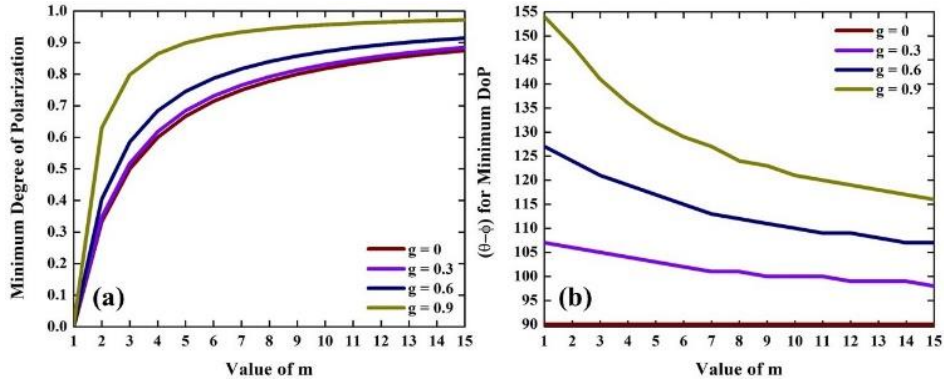

Fig. 4. Change of the (a) magnitude and (b) position of minimum DoP with *m* for different values of *g*.

In case of variation with *m*, the magnitude of the minimum DoP is also found to be increasing monotonically with *m*, as can be seen in Fig. 2, although it is observed from Fig. 4(a) that the nature of the variation is different compared to the variation with *g*. It can be observed from Fig. 4(b) that change in the value of *m* also results in shift in the position of the minimum DoP, except for the case of g = 0, where the position of the minimum DoP is found to be remain unchanged at $90^0$, although the magnitude of the minimum DoP increases. With the increase of *m*, the position of the minimum DoP is observed to be shifting towards $90^0$ from the value at m = 1. The change in the position of minimum DoP with *m* is also found to be affected with the increase of *g*, and increase in the value of *g* is observed to be leading to a sharper variation. The results presented in Figs. 3 and 4 do not include the case of g = 1, as it is found in Fig. 1 that the DoP, in this case, is independent of the polarization diversity of the constituent speckle patterns, and further study also reveals that this variation is not affected by the value of *m*.

In conclusion, a generalized theory of the spatial CP property for superposition of two speckle patterns with polarization diversity is presented in detail, and it has been shown that the polarization diversity combined with the mutual correlation of the two constituent speckle patterns have a tremendous effect on the spatial CP property of the superposed speckle pattern. It is shown that change in the mutual correlation of the two constituent speckle patterns i.e. the value of *g* changes the position of the minimum of the maximum DoC and DoP, whereas change in the average intensity ratio i.e. *m* is observed to be reducing the difference between the maxima and minima of the maximum DoC and

DoP. This study may be useful to understand the effect of superposition of two speckle patterns on the resultant speckle patterns, which can be optimized as per the requirement of the application. Moreover, this study can be extended to the temporal domain as well.